\documentclass[fleqn,10pt]{wlscirep}
\usepackage[utf8]{inputenc}
\usepackage[T1]{fontenc}
\usepackage{array}
\usepackage{color, colortbl}
\title{Identifying Epigenetic Signature of Breast Cancer with Machine Learning}

\author[*]{Maxim Vaysburd}

\affil[*]{Stuyvesant High School, New York, NY, USA}
\affil[*]{mvaysburd10@stuy.edu}


\begin{abstract}
The research reported in this paper identifies the epigenetic biomarker (methylation beta pattern) of breast cancer. Many cancers are triggered by abnormal gene expression levels caused by aberrant methylation of CpG sites in the DNA. In order to develop early diagnostics of cancer-causing methylations and to develop a treatment, it is necessary to identify a few dozen key cancer-related CpG methylation sites out of the millions of locations in the DNA. This research used public TCGA dataset to train a TensorFlow machine learning model to classify breast cancer versus non-breast-cancer tissue samples, based on over 300,000 methylation beta values in each sample. L1 regularization was applied to identify the CpG methylation sites most important for accurate classification. It was hypothesized that CpG sites with the highest learned model weights correspond to DNA locations most relevant to breast cancer. A reduced model trained on methylation betas of just the 25 CpG sites having the highest weights in the full model (trained on methylation betas at over 300,000 CpG sites) has achieved over 94\% accuracy on evaluation data, confirming that the identified 25 CpG sites are indeed a biomarker of breast cancer.
\end{abstract}

\usepackage{url}
\begin{document}

\flushbottom
\maketitle
\thispagestyle{empty}

\section*{Introduction}

Methylation is the addition of methyl groups to nucleotides in the DNA molecule. 
The effects of methylation are particularly important at CpG sites, which are regions of DNA where a cytosine nucleotide is followed by a guanine nucleotide. When a cytosine nucleotide at a CpG site in a promoter region of DNA is methylated, the resulting effect is downregulation of the expression of the respective gene. Methylation can occur at millions of CpG sites throughout the DNA. The CpG sites can be both overmethylated and undermethylated compared to normal levels \cite{Wei:2016,Bernstein:2015}.

Abnormalities in DNA methylation are important causes of development and progression of cancer \cite{Jones:2002,Baylin:2011,Beggs:2013,Bernstein:2015,Wei:2016,Pan:2017}. Indeed, aberrations in methylation levels often appear well before the cancer develops, even while the cells still look completely normal under the microscope \cite{Bernstein:2013,Pan:2017}. 

In cancers, it is common that CpG sites are over-methylated at genes that make proteins promoting tumor suppressor genes \cite{Jones:2002,Li:2016,Pan:2017,Vidal:2017}, DNA repair genes \cite{Bernstein:2013,Jin:2012}, and genes that control cell cycle, differentiation, and apoptosis \cite{Suva:2013}. This results in silencing of tumor suppression and DNA repair, which contributes to development of cancer via an increased occurrence of unrepaired gene mutations and runaway cancer cells \cite{Bernstein:2015}. The silencing of tumor suppressor genes by abnormal over-methylation is in fact much more frequent than their inactivation by DNA mutations \cite{Vogelstein:2013}.  

At the same time, CpG sites in cancer are often under-methylated at oncogenes, which can cause tumors when overexpressed \cite{Pan:2017}. Genome-wide average methylation levels are also significantly lower in tumors than in healthy cells, and are still lower in metastatic cancer cells \cite{Jones:2002,Li:2016,Vidal:2017}.

The presence of both hypo- and hyper-methylation in cancer cells means that effective pharmaceutical treatments will have to be able to correct dangerous methylation levels in both directions: decrease methylation at CpG sites where abnormally high methylation is harmful, and increase at sites where abnormally low methylation is harmful. Methylation abnormalities are reversible \cite{Jones:2002}, and there are indeed pharmaceuticals approved by the FDA that can be used to reduce methylation levels genome-wide \cite{Kantarjian:2006,Kelly:2010,Baylin:2011,Yang:2014,Pan:2017}. They are used to treat precancerous conditions and certain cancers caused by over-methylation of tumor suppressor genes. However, those medications have side effects and may cause other cancers by increasing expression of oncogenes and genes that promote cell invasiveness, and by causing general instability of the genome via broad upregulation of gene expressions\cite{Kelly:2010,Li:2017}.

Several years ago, CRISPR technology was developed to correct genetic abnormalities at precise locations in the DNA \cite{Cong:2013}. It is plausible that in another several years a technology will be developed to correct epigenetic abnormalities at precisely specified CpG methylation sites. In order to be able to apply this technology for prevention and treatment of cancer, it will be necessary to know exactly which CpG methylation sites to correct, and in which direction to change the methylation, out of the hundreds of thousands of CpG sites in the DNA whose methylation can be measured with currently available sequencing technologies.

This research identifies the key CpG sites in the DNA whose methylation levels comprise the biomarker (epigenetic signature) of breast cancer, and which can be used to identify, with high accuracy, breast cancer tissue samples among tissue samples of various other types of cancer. 

\section*{Results and Discussion}

\begin{figure}[ht]
\centering
\includegraphics[width=\linewidth]{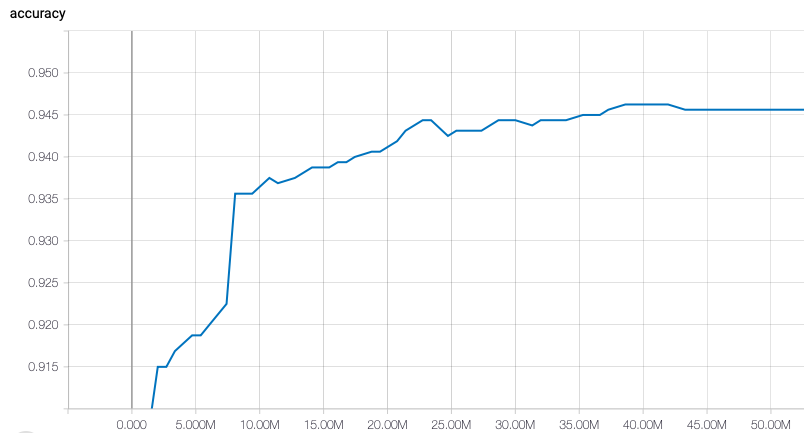}
\caption{Accuracy of the second classification model trained on methylation betas of 25 CpG probes having the highest weights in the first model. Accuracy was computed on the evaluation data (which was not used in training the model). The horizontal axis shows the number of training steps completed by the model. Accuracy increased during the first forty million steps and stabilized around 94.6\%.
}
\label{fig:accuracy_top_25}
\end{figure}

This research has identified 25 CpG locations in the DNA whose methylation beta patterns are sufficient to classify breast cancer vs. non-breast-cancer solid tissue samples with a 94.6\% accuracy using a machine learning classification model trained on methylation beta values at those 25 locations (Figure \ref{fig:accuracy_top_25}). 

\begin{figure}[ht]
\centering
\includegraphics[width=\linewidth]{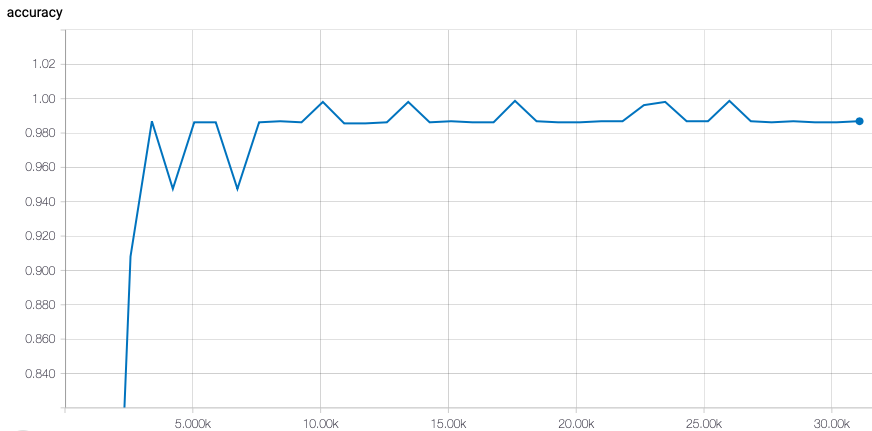}
\caption{Accuracy of the classification model using methylation betas for 323,179 distinct CpG probe IDs. Accuracy was computed on the evaluation data (which was not used in training the model). The horizontal axis shows the number of training steps completed by the model. Accuracy increased during the first several thousand training steps and stabilized around 98.6\%.
}
\label{fig:accuracy_full_model}
\end{figure}

This represents only a modest drop from the 98.6\% accuracy on evaluation data in the breast cancer vs. non-breast-cancer classification model trained on the full set of 323,179 CpG sites  (Figure \ref{fig:accuracy_full_model}). The retention of most of the accuracy indicates that the methylation beta patterns at the identified 25 CpG sites are, indeed, a biomarker of breast cancer and can be used to distinguish it from other types of cancer with high accuracy.

Nine of the genes containing the identified 25 CpG sites ( TAF3, SFRP4, SBDS, PPP2R2A, RASSF4, MTA2, ZNF135, CDHR2, PACS2)  turn out to be directly involved in the cell cycle and to regulate cell growth, differentiation, division, apoptosis, and tumor suppression (Table \ref{tab:gene-function}). These findings further confirm the significance of the CpG methylation sites identified by this research as an epigenetic biomarker of breast cancer.

Several of the genes containing the identified 25 CpG sites have unknown functions as of yet, and could turn out to regulate tumor-related activities as well. Their role in breast cancer is yet to be determined, and presents a direction for future research.

The mean methylation beta values for the 25 CpG sites, for the breast-cancer and non-breast-cancer solid tissue samples, are shown in Table \ref{tab:methylation-means}.

\begin{table}[ht]
\centering
\begin{tabular}{|c|c|c|c|p{72mm}|}
\hline
\textbf{CpG probe ID} & 
\textbf{Chromosome} & 
\textbf{Position} & 
\textbf{Gene symbol} & 
\textbf{Function of the gene} \\
\hline
cg24525395 & chr2 & 60569992 & N/A & To be determined. \\
\hline
cg14371620 & chr2 & 118182593 & AC093901.1 & To be determined. \\
\hline
cg06282247 & chr2 & 150425327 & N/A & To be determined. \\
\hline
cg16171484 & chr2 & 222425796 & SGPP2 & To be determined. \\
\hline
cg00332146 & chr5 & 176548900 & CDHR2 & Involved in tumor suppression, inhibition of cell proliferation. \\
\hline
cg00332146 & chr5 & 176548900 & RN7SL684P & To be determined. \\
\hline
cg27064266 & chr6 & 27865624 & HIST1H2AL & To be determined. \\
\hline
cg27064266 & chr6 & 27865624 & HIST1H2BN & To be determined. \\
\hline
cg27064266 & chr6 & 27865624 & HIST1H2BPS2 & To be determined. \\
\hline
cg27064266 & chr6 & 27865624 & Z98744.2 & To be determined. \\
\hline
cg02335619 & chr6 & 31835168 & C6orf48 & To be determined. \\
\hline
cg02335619 & chr6 & 31835168 & SNORD48 & To be determined. \\
\hline
cg06962768 & chr7 & 37920180 & EPDR1 & To be determined. \\
\hline
cg06962768 & chr7 & 37920180 & SFRP4 & Involved in apoptosis and regulation of cell growth and differentiation. \\
\hline
cg00498438 & chr7 & 66995689 & SBDS & Affects apoptosis. \\
\hline
cg00498438 & chr7 & 66995689 & TYW1 & To be determined. \\
\hline
cg22274662 & chr7 & 149497004 & ZNF746 & To be determined. \\
\hline
cg13877285 & chr8 & 26291357 & PPP2R2A & Negative control of cell growth and division. \\
\hline
cg04693895 & chr10 & 385155 & DIP2C & To be determined. \\
\hline
cg12777293 & chr10 & 7818591 & TAF3 & Acts as an antiapoptotic factor. \\
\hline
cg27139956 & chr10 & 44974842 & C10orf10 & To be determined. \\
\hline
cg27139956 & chr10 & 44974842 & RASSF4 &
Tumor suppression. Promotes apoptosis and cell cycle arrest. Inhibits tumor cell growth and colony formation. Broadly expressed in normal cells, downregulated by methylation in tumor cells. \\
\hline
cg13298116 & chr11 & 62602387 & EML3 & To be determined.\\
\hline
cg13298116 & chr11 & 62602387 & MTA2 &
Associated with metastasis of cancer cells. Regulates gene expression. Is overexpressed in cancers. Correlates with cancer invasiveness and aggressiveness. Overexpression of MTA2 promotes metastasis of breast cancer cells. Regulates pathways involved in regulation, apoptosis, growth of normal and cancer cells. \\
\hline
cg14100748 & chr14 & 90577405 & TTC7B & To be determined. \\
\hline
cg12158535 & chr14 & 105390913 & PACS2 & Involved in cell apoptosis. \\
\hline
cg01027010 & chr16 & 31565018 & CTD-2014E2.5 & To be determined. \\
\hline
cg00103783 & chr17 & 7583931 & AC113189.5 & To be determined.\\
\hline
cg00103783 & chr17 & 7583931 & MPDU1 & To be determined.\\
\hline
cg14204735 & chr17 & 63446943 & CYB561 & To be determined.\\
\hline
cg15751406 & chr19 & 8823867 & CTD-2529P6.3 & To be determined.\\
\hline
cg15751406 & chr19 & 8823867 & ZNF558 & To be determined.\\
\hline
cg09907936 & chr19 & 58059098 & ZNF135 &
Involved in both normal and abnormal cellular 
proliferation and differentiation. \\
\hline
cg23060618 & chr21 & 43301197 & N/A & To be determined.\\
\hline
cg03113871 & chr22 & 30921800 & MORC2-AS1 & To be determined.\\
\hline
cg20180585 & chr22 & 41381751 & TEF & To be determined.\\
\hline
\end{tabular}
\caption{\label{tab:gene-function}Mapping of the top 25 CpG sites identified by the model to chromosome positions, genes, and respective gene functions, based on the GeneCards and OMIM databases. Nine of the genes are directly involved in the cell cycle and regulation of cell growth, differentiation, division, apoptosis, and tumor suppression. The function of the remaining genes is yet to be determined.
}
\end{table}

\begin{table}[ht]
\centering
\begin{tabular}{|c|c|c|}
\hline
\textbf{CpG probe ID} & 
\textbf{non-BRCA beta mean} & 
\textbf{BRCA beta mean}  \\
\hline
cg03113871 & 0.146 & 0.105 \\ \hline
cg24525395 & 0.153 & 0.161 \\ \hline
cg16171484 & 0.542 & 0.734 \\ \hline
cg14100748 & 0.888 & 0.899 \\ \hline
cg00103783 & 0.024 & 0.022 \\ \hline
cg22274662 & 0.042 & 0.042 \\ \hline
cg20180585 & 0.013 & 0.012 \\ \hline
cg27139956 & 0.158 & 0.09 \\ \hline
cg13877285 & 0.029 & 0.022 \\ \hline
cg04693895 & 0.931 & 0.958 \\ \hline
cg23060618 & 0.742 & 0.778 \\ \hline
cg12158535 & 0.682 & 0.667 \\ \hline
cg00498438 & 0.019 & 0.019 \\ \hline
cg15751406 & 0.93 & 0.95 \\ \hline
cg01027010 & 0.604 & 0.447 \\ \hline
cg02335619 & 0.039 & 0.036 \\ \hline
cg00332146 & 0.576 & 0.699 \\ \hline
cg14204735 & 0.204 & 0.053 \\ \hline
cg12777293 & 0.014 & 0.013 \\ \hline
cg13298116 & 0.143 & 0.092 \\ \hline
cg06962768 & 0.759 & 0.695 \\ \hline
cg27064266 & 0.132 & 0.122 \\ \hline
cg14371620 & 0.785 & 0.783 \\ \hline
cg09907936 & 0.344 & 0.191 \\ \hline
cg06282247 & 0.48 & 0.625 \\ \hline
\end{tabular}
\caption{\label{tab:methylation-means}Mean methylation betas for the top 25 model-selected CpG probe IDs, for the BRCA (breast cancer) and non-BRCA solid tissue samples.}
\end{table}

The methodology applied in this research can be extended to identify epigenetic biomarkers for types of cancer other than breast cancer (the BRCA study in TCGA). Another direction for future research is to learn epigenetic biomarkers shared by multiple cancer types, by training a machine learning model to classify cancerous vs. normal tissues.

The knowledge of the epigenetic signature of breast cancer identified in this research is valuable in several ways: it can enable development of medications to correct abnormal methylation at precisely the CpG sites that control expression of genes involved in the development and progression of cancer, and it can also be used for early diagnostics and prevention of breast cancer, at the stage where analysis of tissue cells under the microscope (in biopsies) doesn't show any abnormalities, yet the epigenetic signature of cancer is already present.

\section*{Materials and Methods}

The research reported in this paper was performed using data in the public DNA methylation betas dataset\cite{TCGA-data:2018} produced by The Cancer Genome Atlas Program\cite{TCGA-program:2018}, containing tissue samples from 32 different types of cancer. The data has been retrieved from the public dataset \href{https://bigquery.cloud.google.com/dataset/isb-cgc:TCGA_hg38_data_v0}{isb-cgc:TCGA\_hg38\_data\_v0}, hosted by the Institute for Systems Biology - Cancer Genomics Cloud (ISB-CGC)\cite{CancerGenomicsCloud} project on Google Cloud\cite{google-cloud}. The data has been accessed via Google BigQuery\cite{bigquery}. Methylation betas for more than 400,000 CpG sites in 8,703 solid cancer tissue samples (identified by unique aliquot barcodes) were retrieved from the \href{https://bigquery.cloud.google.com/table/isb-cgc:TCGA_hg38_data_v0.DNA_Methylation}{TCGA\_hg38\_data\_v0.DNA\_Methylation} table. 

A subset of samples and CpG probes were selected for use in the project, such that each included sample would contain methylation betas for the same set of CpG probes. This step reduced the size of the research dataset to 8,126 aliquots, with the same 323,179 probes included with each aliquot.

In the next step, training examples were constructed for the machine learning classification model, with each training example comprising a list of 323,179 methylation beta values for one aliquot, as the input features of the model. For each training example, the label was set to 1 for aliquots derived from BRCA (breast cancer study) tissue samples, and 0 for aliquots from all other included TCGA cancer studies. In order to prevent the classification model from being biased towards 1 or 0 values of the label, the same number of training examples were constructed for BRCA as for all other (non-BRCA) studies combined. Within the non-BRCA class (comprising training examples with the label 0), the same number of training examples were constructed from aliquots in each included non-BRCA study. The number of solid cancer tissue samples (aliquots) available for each cancer study in the TCGA dataset ranges from 763 for BRCA (breast cancer study) to 10 for OV. Altogether, training examples were generated using aliquots from the 22 TCGA studies having the highest number of available tissue samples.

\begin{figure}[ht]
\centering
\includegraphics[width=\linewidth]{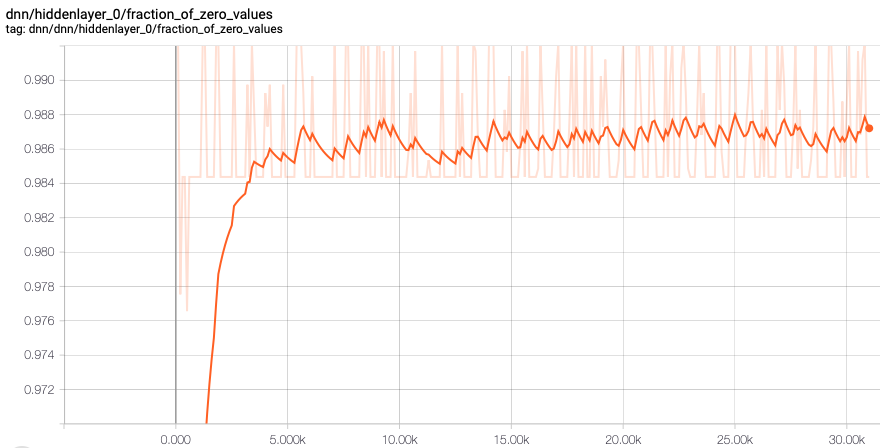}
\caption{Fraction of zero weights in the classification model using methylation betas for 323,179 distinct CpG probe IDs. The horizontal axis shows the number of training steps completed by the model.
}
\label{fig:fraction_of_zero_weights}
\end{figure}

In the following step, a classification model was trained using the \href{https://www.tensorflow.org/api_docs/python/tf/estimator/DNNClassifier}{tf.estimator.DNNClassifier} class available in the TensorFlow\cite{tensorflow} machine learning library. The classifier has been configured with one fully connected layer containing 128 hidden units, and two classes (1 or 0) corresponding to training examples as being either BRCA or non-BRCA. The Proximal Adagrad optimizer (available as the \href{https://www.tensorflow.org/api_docs/python/tf/train/ProximalAdagradOptimizer}{tf.train.ProximalAdagradOptimizer} class in TensorFlow) was used to train the model with L1 regularization enabled. The regularization level was set to 0.005 to force the model to use as few input features as possible without significantly hurting the accuracy of classification. The goal in using L1 regularization was to identify CpG sites in the genome whose methylation beta values are the most indispensable for the model for accurate classification of training examples. The model was trained using 80\% of all training examples constructed in the previous step, with the remaining 20\% of examples held out for the evaluation of results. Upon completion of 30,000 training steps, 98.7\% of weights in the model’s fully connected layer had the value of zero (Figure \ref{fig:fraction_of_zero_weights}).

\begin{figure}[ht]
\centering
\includegraphics[width=\linewidth]{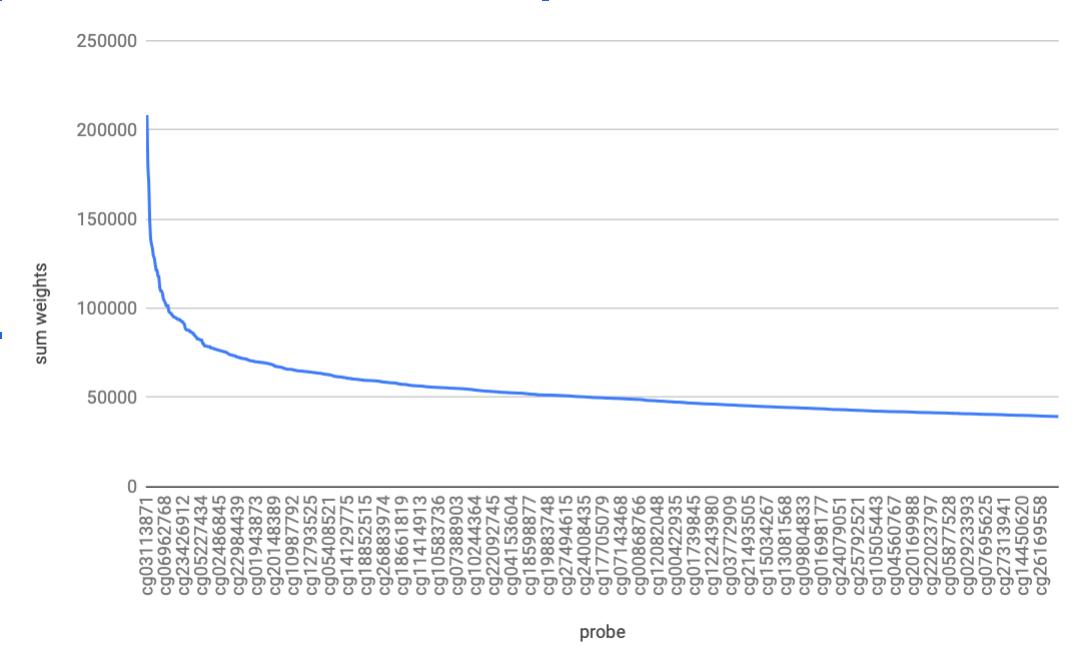}
\caption{Sum of the absolute values of model weights for each CpG probe ID. The chart includes the top 1,000 probe IDs (out of 323,179 probe IDs used to train the first classification model), ordered by the sum of absolute values of weights, in descending order. The weights of the top 25 CpG sites are much higher than the weights of the remaining probes.
}
\label{fig:sum_weights}
\end{figure}

In the next step, the CpG sites (probe IDs) were sorted by the sums of absolute values of their weights in the classification model’s fully connected layer (Figure \ref{fig:sum_weights}), and the 25 CpG probe IDs with the highest model weights were selected. 

A new classification model was then trained using the same 80\% of training examples as in the original model but with each example constrained to include methylation betas of the selected top 25 probe IDs only (instead of the 323,179 betas used to train the first model). The same 20\% of examples (constrained to the top 25 probe IDs) were used for the evaluation of the model. The training of the second model was configured with the same settings as in the first model, except that this time regularization was not applied. The purpose of the second model was to verify that the selected top 25 probes are indeed the key CpG sites comprising the biomarker of breast cancer.

Upon completion of model training, the data in the
\href{https://bigquery.cloud.google.com/table/isb-cgc:platform_reference.methylation_annotation}{isb-cgc:platform\_reference.GDC\_hg38\_methylation\_annotation} table in BigQuery was used to identify the genes containing the model-selected top 25 CpG sites. The GeneCards\cite{GeneCards:2016} and OMIM\cite{Hamosh:2005} databases were then used to look up the functions of those genes. 

Finally, the average methylation beta values were computed for the selected top 25 CpG sites, for BRCA and non-BRCA tissue samples in TCGA.

\section*{Data Availability}

The Cancer Genome Atlas data analysed during the current study is available in the
public BigQuery dataset repository \href{https://bigquery.cloud.google.com/dataset/isb-cgc:TCGA_hg38_data_v0}{isb-cgc:TCGA\_hg38\_data\_v0} in Google Cloud, at 
\url{https://bigquery.cloud.google.com/dataset/isb-cgc:TCGA_hg38_data_v0}.

\bibliography{references}

\section*{Author contributions statement}

M.V. conceived and conducted the research reported in this paper, and analyzed the results. 

\section*{Acknowledgements}

The results attained in this research are based upon data generated by the TCGA Research Network: \url{https://www.cancer.gov/tcga}.

\section*{Additional information}

The author declares no competing interests.

\end{document}